\documentclass[aps, prb, twocolumn, amssymb, amsmath, superscriptaddress]{revtex4-1}

\usepackage{bm}
\usepackage{times}
\usepackage{graphicx}
\usepackage[colorlinks,citecolor=blue,linkcolor=blue]{hyperref}
\usepackage{array}
\usepackage{float}
\usepackage{colortbl}
\usepackage{multirow}
\usepackage{tabularx}

\begin{document}

\title{Large thermoelectric power factor of high-mobility 1\textbf{\emph{T$''$}} phase of transition-metal dichalcogenides}

\author{Yanfeng Ge}
\affiliation{State Key Laboratory of Metastable Materials Science and Technology \& Key Laboratory for Microstructural Material Physics of Hebei Province, School of Science, Yanshan University, Qinhuangdao, 066004, China}

\author{Wenhui Wan}
\affiliation{State Key Laboratory of Metastable Materials Science and Technology \& Key Laboratory for Microstructural Material Physics of Hebei Province, School of Science, Yanshan University, Qinhuangdao, 066004, China}

\author{Yulu Ren}
\affiliation{State Key Laboratory of Metastable Materials Science and Technology \& Key Laboratory for Microstructural Material Physics of Hebei Province, School of Science, Yanshan University, Qinhuangdao, 066004, China}

\author{Yong Liu}\email{yongliu@ysu.edu.cn}
\affiliation{State Key Laboratory of Metastable Materials Science and Technology \& Key Laboratory for Microstructural Material Physics of Hebei Province, School of Science, Yanshan University, Qinhuangdao, 066004, China}
\date{\today}

\begin{abstract}
The experimental studies about monolayer transition metal dichalcogenides in the recent year reveal this kind of compounds have many metastable phases with unique physical properties, not just 1H phases. Here, we focus on the 1$T''$ phase and systematically investigate
the electronic structures and transport properties of MX$_2$ (M=Mo, W; X=S, Se, Te) using the first-principles calculations with Boltzmann transport theory. And among them, only three molybdenum compounds has small direct bandgap at K point, which derive from the distortion of octahedral-coordination [MoX$_6$]. For these three cases, hole carrier mobility of MoSe$_2$ is estimated as 690 cm$^2$/Vs at room temperature, far more high than that of other two MoX$_2$. For the reason, the combination of the modest carrier effective mass and weak electron-phonon coupling lead to the outstanding transport performance of MoSe$_2$. The Seebeck coefficient of MoSe$_2$ is also evaluated as high as $\sim$ 300 $\mu$V/K at room temperature. Due to the temperature dependent mobility of T$^{-1.9}$ and higher Seebeck coefficient at low temperature, it is found that MoSe$_2$ has a large thermoelectric power factor around 6 10$^{-3}$ W/mK$^2$ in the low to intermediate temperature range. The present results suggests 1$T''$ MoSe$_2$ maybe a excellent candidate for thermoelectric material.
\end{abstract}

\maketitle

\section{Introduction}

Since the discovery of 2D materials, transition metal dichalcogenides (TMDCs)~\cite{Wang2012,Feng2012,Conley2013,Castellanos-Gomez2013,Chhowalla2013}
are particularly interesting due to the semiconducting characteristics with strong stability and large flexibility.
The unique structural, mechanical, optical, electrical, and thermal properties~\cite{Fiori2014,Bernardi2017} make they have potential applications in photovoltaics~\cite{Britnell2013}, transistors~\cite{Fang2012,Liu2013,Yoon2011,Mak2013,Qiu2015}, Optoelectronic~\cite{Huo2014}, photodetector and molecular sensing~\cite{Wang2012}.
The usual crystallographic form of monolayer TMDCs is the hexagonal 1H phase, in which TMDCs have the high on-off ratio (10$^8$) with
carrier mobility of 200 cm$^2$/Vs at room temperature~\cite{Radisavljevic2011,Levi2013}.
The exciton energy and strong spin-valley coupling of 1H phase also provide novel platform for intriguing nanoelectronic devices~\cite{Xu2014,Mak2014,Cui2015,Roy2013,Qi2015}.
TMDCs also exhibit other metastable trigonal polymorphic forms with different degrees of structural distortion (1T, 1$T'$, 1$T''$, and 1$T'''$)~\cite{Song2015,Zhang2018,Linghu2019,Zhao2018,Kan2014,Calandra2013,Bruyer2016,Zhuang2017,Pal2017,Zhou2018,Singh2015}.
Thereinto 1T is the primary structure and adopts the octahedral coordination with point group D$_{3d}$. In the octahedral crystal field, 4d orbitals of Mo atoms are split into the e$_{\rm g}$ orbitals (d$_{x^2-y^2}$, d$_{z^2}$) over t$_{\rm 2g}$ orbitals (d$_{xy}$, d$_{xz}$, d$_{yz}$), and the partially filled t$_{\rm 2g}$ orbitals induce metallic conductivity~\cite{Chhowalla2013}.
Due to the Peierls instability, the distortion of octahedral [MoS$_6$] in 1T phase can result in other metastable polymorphs with lower symmetry~\cite{Chou2015,Eda2012}, where Mo-Mo associations take place, such as the dimerization (1$T'$)~\cite{Yu2018} and trimerization (1$T''$ and 1$T'''$)~\cite{Fang2018,Shirodkar2014,Shang2018}.
The spontaneous symmetry breaking of structural distortion lift the degeneracy of electronic states to lower the energy.
In these metastable structures, the higher conductivity of metallic 1T phase make them as excellent electrocatalysts for hydrogen evolution, rechargeable batteries and supercapacitors~\cite{Acerce2015,Voiry2013,Lukowski2013}.
Many novel physical properties are also revealed in the distorted structures. For example, the strong spin-orbital coupling (SOC) make 1$T'$ TMDCs to be large-gap quantum spin hall insulators~\cite{Qian2014,Choe2016,Tang2017}.
The nontrivial geometry with the trimerization of Mo atoms in 1$T'''$ phase lead to the ferroelectricity with high carrier mobility simultaneously~\cite{Bruyer2016}.

In addition, due to the proportional relation between Seebeck coefficient and the energy derivative of the electronic density of states around Fermi level in the Mott formula~\cite{Cutler1969}, low-dimensional materials TMDCs have natural advantage in thermoelectric (TE) applications, an important and meaningful crossing field of physics, materials and energy~\cite{Bell2008,Heremans2013,Dresselhaus2007,Zhao2014}.
Therefore, more recently people have paid attention to TMDCs in the prospect of thermoelectricity~\cite{Huang2013,Fan2014,Huang2014,Babaei2014,Wang2017,Wu2014,Yoshida2016,TWang2016}.
The efficiency of TE materials depends on their dimensionless figure-of-merit ZT defined as ZT = $\sigma$S$^2$T/$\kappa$. S is the Seebeck coefficient, $\sigma$ is the electrical conductivity, T is the absolute temperature, and $\kappa$ is the total thermal conductivity and characterizes the heat leakage.
Reaching high ZT has remained demanding because of the complicated relation between these individual parameters, especially the electrical conductivity and Seebeck coefficient. In general, the competition appears between these two properties, a small carrier effective mass favors high $\sigma$, but opposes a large S. Hence, power factor ($\sigma$S$^2$) is often used to represent the electron energy
conversion capability in TE materials.
Recently, by using electric double-layer technique (EDLT) with the gate dielectrics of ionic liquids, researchers measure the ultrathin WSe$_2$ single crystals and obtain an power factor of $\sim$ 4 10$^{-3}$ W/mK$^2$ ~\cite{Yoshida2016}.
Another experiment report a power factor of MoS$_2$ as large as 8.5 10$^{-3}$ W/mK$^2$ at room temperature~\cite{Hippalgaonkar2017}, exfoliated samples by the scotch-tape method. Moveover, it is found that the Kondo effect can improve the power factor of MoS$_2$~\cite{Wu2019} to much quite high value of 50 10$^{-3}$ W/mK$^2$. While the power factor in other TE experiment about TMDCs is much lower than that in the above experiments. The main reasons is that low electrical conductivity limits the power factor for TE applications.

As is well-known, prevalent TE materials are heavily-doped small-bandgap semiconductors~\cite{Gascoin2005,Lee2012,Goldsmid2014}, which can hold the balance between high Seebeck coefficient of semiconductor and high electrical conductivity of metals.
Therefore, in the present work, we focus on the 1$T''$ phase of transition-metal dichalcogenides with small bandgap, such as 0.1 eV in 1$T''$-MoS$_2$~\cite{Zhuang2017}, and explains the origin of small bandgap from the structure distortion. Since carrier doping at high concentration of EDLT has been successfully used to improve the performance of TMDCs, this work also systematically explore the dependents of electronic transport for a large range of carrier-doping concentrations by considering the electron-phonon coupling. The lower carrier effective mass and the weakest electron-phonon scattering make 1$T''$ MoSe$_2$ has high mobility of 690 cm$^2$/Vs at room temperature. Moreover, duo to the advantages of small bandgap and suitable carrier effective mass on the enhancement of Seebeck coefficient (300 $\mu$V/K), we obtain that MoSe$_2$ has high value around 6 10$^{-3}$ W/mK$^2$ in a larger temperature range.

\section{Methods}

In the diffusive transport regime, electronic transport of a material can be calculated based on the Boltzmann transport equation (BTE).
In the consideration of electron-phonon scattering in and out of the state $|n \mathbf{k}\rangle$ ($\varepsilon_{n \mathbf{k}}$), via emission or absorption of phonons ($\omega_{\mathbf{q} \nu}$), the relaxation time $\tau_{n \mathbf{k}}^{0}$ is associated with the imaginary part of the Fan-Migdal electron self-energy~\cite{Giustino2017}, defined by~\cite{Ponce2018}
\begin{equation}
\begin{aligned}
 \frac{1}{\tau_{n \mathbf{k}}^{0}}=& \frac{2 \pi}{\hbar} \sum_{m v} \int \frac{d \mathbf{q}}{\Omega_{\mathrm{BZ}}}\left|g_{m n v}(\mathbf{k}, \mathbf{q})\right|^{2} \\ & \times[\left(1-f_{m \mathbf{k}+\mathbf{q}}^{0}+n_{\mathbf{q} v}\right) \delta\left(\varepsilon_{n \mathbf{k}}-\varepsilon_{m \mathbf{k}+\mathbf{q}}-\hbar \omega_{\mathbf{q} v}\right).\\ &+\left(f_{m \mathbf{k}+\mathbf{q}}^{0}+n_{\mathbf{q} v}\right) \delta\left(\varepsilon_{n \mathbf{k}}-\varepsilon_{m \mathbf{k}+\mathbf{q}}+\hbar \omega_{\mathbf{q} \nu}\right)],
 \end{aligned}
 \label{eq:time}
\end{equation}
where $\Omega_{\rm BZ}$ is the volume of the first Brillouin zone,
$f$ and $n$ are the Fermi-Dirac and Bose-Einstein distribution functions, respectively.
In Eq.(\ref{eq:time}),
The electron-phonon matrix elements $g_{m n v}(\mathbf{k}, \mathbf{q})$ are the probability
amplitude for scattering from an initial electronic state $|n \mathbf{k}\rangle$
into a final state $|m \mathbf{k}+\mathbf{q}\rangle$ via a phonon $|\mathbf{q} \nu \rangle$, as obtained from density-functional perturbation theory (DFPT)~\cite{Baroni2001,Giustino2017,Ponce2018}.

In the self-energy relaxation time approximation (SERTA)~\cite{Ponce2018}, the electron carrier mobility takes the simple form
\begin{equation}
\mu_{\mathrm{e}}=\frac{-e}{n_{\mathrm{e}} \Omega} \sum_{n \in \mathrm{CB}} \int \frac{d \mathbf{k}}{\Omega_{\mathrm{BZ}}} \frac{\partial f_{n \mathrm{k}}^{0}}{\partial \varepsilon_{n \mathrm{k}}} v_{n \mathbf{k}} v_{n \mathbf{k}} \tau_{n \mathbf{k}}^{0},
 \label{eq:mu}
\end{equation}
where $v_{n \mathbf{k}}$ is the group velocity of electronic state $|n \mathbf{k}\rangle$ and $\Omega$ is the volume of the crystalline unit cell.
Based on the relaxation time $\tau_{n \mathbf{k}}^{0}$, the TE transport ($\sigma$ and S) as a function of the chemical potential $\mu$ and of the temperature T is the following expressions~\cite{Huang2018,Pizzi2014}:

\begin{equation}
{\sigma=e^{2} \int \Xi(\varepsilon)\left(-\frac{\partial f^{0}}{\partial \varepsilon}\right) \mathrm{d} \varepsilon},
 \label{eq:sigma}
\end{equation}

\begin{equation}
{S=\frac{e }{\sigma} \int \Xi(\varepsilon)\left(-\frac{\partial f^{0}}{\partial \varepsilon}\right) \frac{\varepsilon-\mu}{ T} \mathrm{d} \varepsilon},
 \label{eq:seeb}
\end{equation}
where $\Xi(\varepsilon)$ is the transport distribution function, defined as $\Xi(\varepsilon)=\sum_{n, \mathbf{k}} v_{n \mathbf{k}} v_{n \mathbf{k}} \tau^0_{n \mathbf{k}} \delta\left(\varepsilon-\varepsilon_{n \mathbf{k}}\right)/\Omega$.

\begin{figure}[ht!]
\centerline{\includegraphics[width=0.45\textwidth]{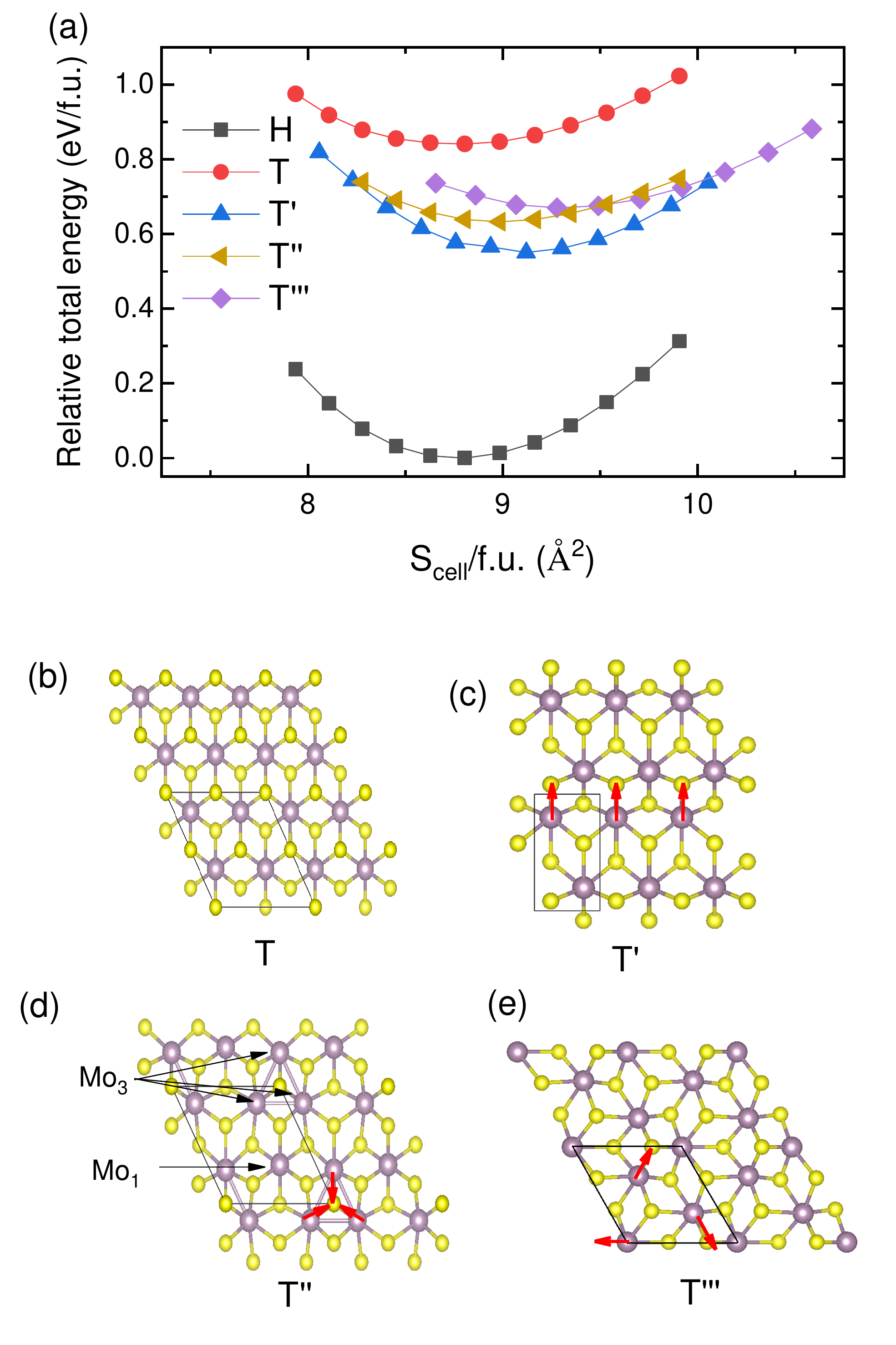}}
\caption{(a) Relative total energy of different monolayer structures with respect to 1H phase. Crystal structure diagrams of 1T (b), 1$T'$ (c), 1$T''$ (d) and 1$T'''$ (e). The black box shows the periodic repeated unit in this work. The distorted 1$T'$, 1$T''$ and 1$T'''$ correspond to dimerization, trimerization and trimerization of Mo atoms, respectively. In 1$T''$ phase, Mo atoms are divided into two categories: trimeric Mo atoms (Mo$_3$) and other Mo atom (Mo$_1$).
\label{fig:struc}}
\end{figure}

\begin{figure*}[t!]
\centerline{\includegraphics[width=1.0\textwidth]{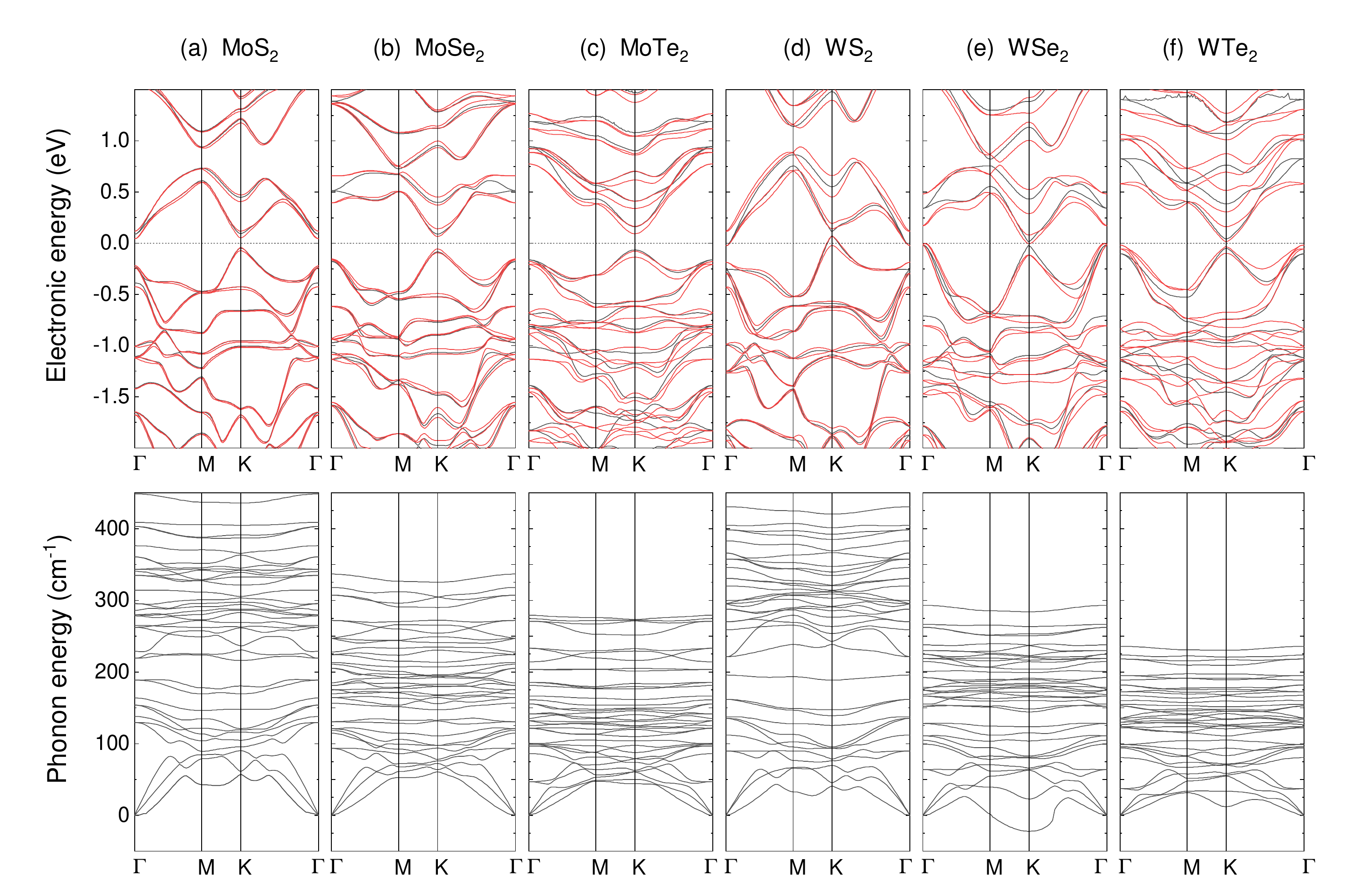}}
\caption{ Band structures (upper half) and phonon spectra (bottom half) of MoS$_2$ (a), MoSe$_2$ (b),  MoTe$_2$ (c),  WS$_2$ (d),  WSe$_2$ (e) and WTe$_2$ (f) in the 1$T''$ phase. Three molybdenum compounds have small direct bandgap at K point. And only WSe$_2$ has imaginary frequency in the phonon spectrum.
\label{fig:bandphonon}}
\end{figure*}

Technical details of the calculations are as follows. All calculations in this work were carried out in the framework of
density-functional theory (DFT) as implemented in the QUANTUM ESPRESSO package~\cite{Giannozzi2017}. The exchange and correlation energy was in the form of Perdew-Burke-Ernzerhof (PBE)~\cite{Perdew1996}. Due to the existence of heavy transition metal element, the fully relativistic SOC was included in all calculations. By requiring convergence of results, the kinetic-energy cutoff of 40 Ry and the Monkhorst-Pack k-mesh of 16$\times$16$\times$1 were used in the calculations dealing with the electronic ground-state properties. The phonon spectra were calculated on a 4$\times$4$\times$1 q grid using DFPT. In order to obtain the stable structure, the atomic positions were relaxed fully with the energy convergence criteria of 10$^{-5}$ Ry and the force convergence criteria of 10$^{-4}$ Ry/a.u. In the monolayer structure, a vacuum layer with 15 \AA\ was set to avoid the interactions between the adjacent atomic layers. Within the EPW code~\cite{Ponce2016} of QUANTUM ESPRESSO in conjunction with the WANNIER90~\cite{Mostofi2008,Mostofi2014}, electron-phonon coupling was calculated on a 40$\times$40$\times$1 q grid with dense k points of 160$\times$160$\times$1 by the Wannier-Fourier interpolation technique of maximally localized Wannier functions~\cite{Giustino2007,Marzari2012}.

\section{Results}

According to the sample preparation in the present experiment~\cite{Song2015,Zhang2018,Linghu2019,Zhao2018,Kan2014,Calandra2013,Bruyer2016,Zhuang2017,Pal2017,Zhou2018,Singh2015}, there are mainly three distorted phases from 1T phase (space group P-3m1). They all have lower symmetry than 1T and can be classified into two cases: dimeric structure 1$T'$ (space group P21/m) and trimeric structure 1$T''$ (space group P3) and 1$T'''$ (space group P31m), as show in Fig.~\ref{fig:struc}. The Peiels distortions of the prototypical 1T phase in the one direction and two directions along lattice vectors lead to the dimerization (1$T'$) and trimerization (1$T''$) of nearest-neighboring transition metal atoms~\cite{Linghu2019}, respectively. And the K$_3$ distortion~\cite{Shirodkar2014}, a small rotary polymerization of three nearest-neighboring Mo atoms, leads to a lower symmetry cell tripled $T'''$ structure. A case study of MoS$_2$, the total energy difference relative to the 1H phase shows that 1T and 1$T'$ phases have the highest and lowest total energy in the metastable phases, respectively, when two trimeric structures have similar total energy. In the 1$T''$ phase, it is found that interatomic distance (2.77 {\AA}) of three Mo atoms in $2a \times 2a$ superstructure is much shorter than that in 1T phase (3.22 {\AA}), marked by Mo$_3$ for simplicity. Other one Mo atom (marked by Mo$_1$) has little deviation relative to the corresponding Mo atom in 1T phase. And the equilibrium lattice constant ($a_0$=6.44 {\AA}) of MoS$_2$ agrees well with the previous studies~\cite{Zhuang2017,Linghu2019}. The heavy chalcogens elongate $a_0$ significantly, accompanied by the slight bigger space between X atomic layer and Mo atomic layer, because of the increase of ionic radius with the atomic number of chalcogens. However, the ionic radius of Mo$^{2+}$ is almost identical to W$^{2+}$, thus the change of $a_0$ induced by the W element is much smaller, as summarized in Tab.~\ref{tab:bandgap}.

\begin{table}[htp!]
\caption{Equilibrium lattice constant ($a_0$), bandgap (E$_{\rm gap}$) and carrier effective mass (m$^*$) of MoS$_2$, MoSe$_2$ and MoTe$_2$.}
\begin{tabular*}{8cm}{@{\extracolsep{\fill}}ccccccccc}
\hline\hline
        &   MoS$_2$ &  MoSe$_2$ & MoTe$_2$ \\
\hline   $a_0$ ({\AA})    &  6.44 &  6.68  & 7.11   \\
         E$_{\rm gap}$ (eV)    &  0.10 &  0.12 & 0.16  \\
         m$^*_{h}$  (m$_0$)  &  0.227 & 0.535   & 0.707 \\
         m$^*_{e}$ (m$_0$) &  0.303 & 0.937   & 0.872  \\
\hline
        &   WS$_2$ &  WSe$_2$ & WTe$_2$ \\
\hline   $a_0$ ({\AA})    &  6.50 &  6.71  & 7.13   \\
         E$_{\rm gap}$ (eV)    &  metal &  metal  & 0.03  \\
         m$^*_{h}$  (m$_0$)  &  - &  -  & 0.566  \\
         m$^*_{e}$ (m$_0$) & -  &  -  &  0.437  \\
\hline \hline
\end{tabular*}
\label{tab:bandgap}
\end{table}

\begin{figure}[ht!]
\centerline{\includegraphics[width=0.45\textwidth]{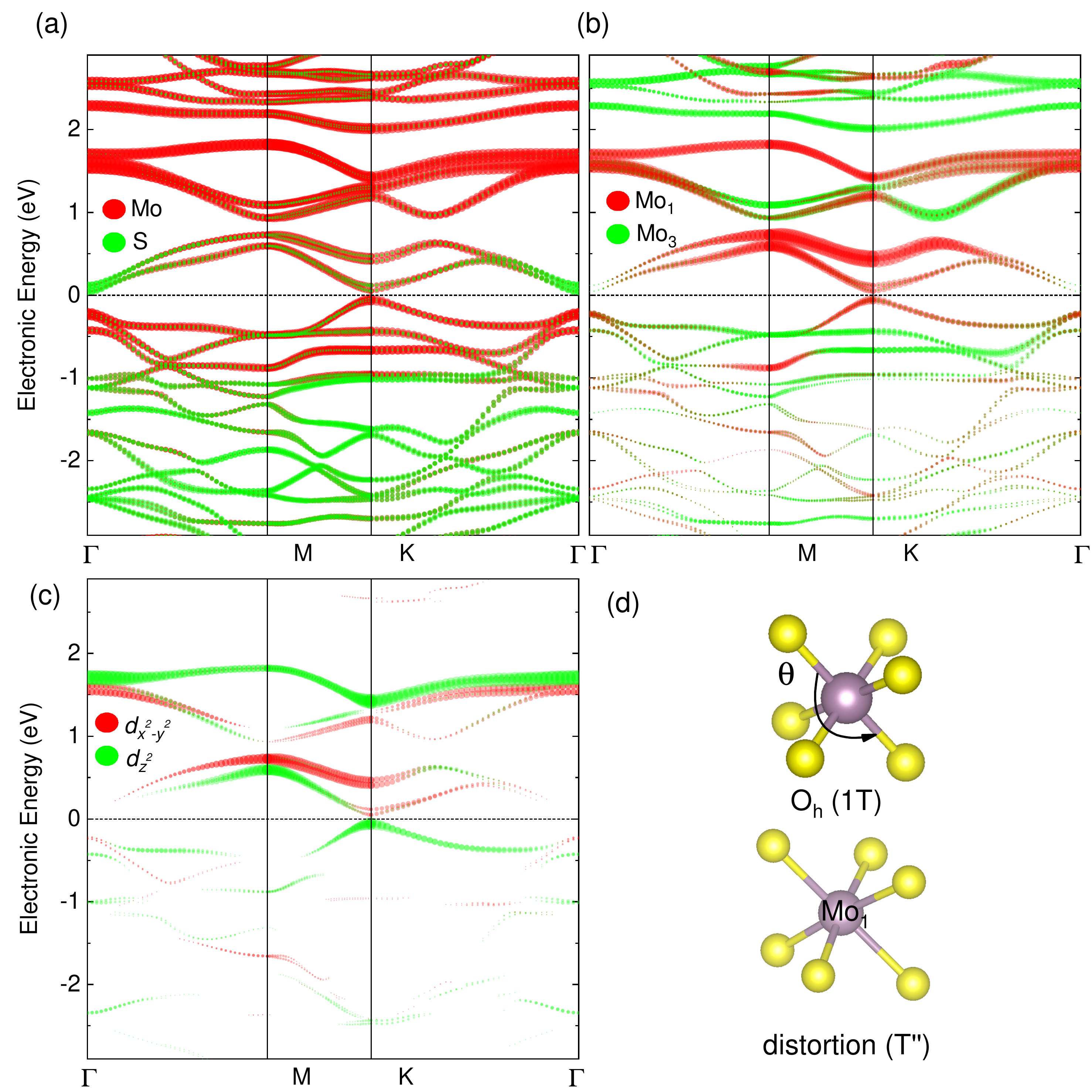}}
\caption{(a) Projected band structure of Mo and S atoms. (b) Projected band structure of different Mo atoms. (c) Projected band structure of different orbits of Mo$_1$ atom. (d) The building of [MoS$_6$] in 1T and 1$T''$ MoS$_2$. The angles between para-position Mo-S bonds ($\theta$) are 180$^{\circ}$ and 175$^{\circ}$ in 1T and 1$T''$ phases, respectively.
\label{fig:projectedband}}
\end{figure}

Because the small bandgap~\cite{Zhao2018} of MoS$_2$ in 1$T''$ phase is advantageous to enhance the thermoelectricity, here we mainly study the 1$T''$-phase MX$_2$ (M=Mo, W; X=S, Se, Te). As shown in Fig.~\ref{fig:bandphonon}, The band structures indicate that three MoX$_2$ all have direct bandgap at K point, when only WTe$_2$ in WX$_2$ is semiconductor with very small indirect-bandgap of 0.03 eV [Tab.~\ref{tab:bandgap}]. For the case of valence band, there is a second energy maximum ($\Gamma_v$) at $\Gamma$ point for all MX$_2$, closing to valence band maximum (VBM) with small energy difference. And the second energy minima ($\Gamma_c$) of conduction band at $\Gamma$ point exists only in MoS$_2$ and WS$_2$. Moreover, the stronger SOC of heavy transition metal atom also generate the larger spin splitting at both conduction band minimum (CBM) and VBM.
Now we analyze the source for the bandgap in 1$T''$ phase according to the projected band structure of MoS$_2$ [Fig.~\ref{fig:projectedband}]. CBM and VBM around the direct bandgap are mainly composed of Mo atomic orbitals, when there are only S atomic orbitals (p$_x$ and p$_y$) at $\Gamma_c$ [Fig.~\ref{fig:projectedband}(a)].
And the short interatomic distance of trimeric Mo$_3$ results in the short Mo-S bonding length as well as the large energy difference between the bonding states and antibonding states of Mo$_3$-4d orbitals. Therefore the distribution of Mo$_3$-4d orbitals are far away from CBM and VBM, which are contribution from the Mo$_1$-d orbitals, as shown in Fig.~\ref{fig:projectedband}(b).
According to the comparation between the distorted octahedral [Mo$_1$S$_6$] and O$_h$-[MoS$_6$] in 1T phase [Fig.~\ref{fig:projectedband}(d)], it is found that the angles between para-position Mo-S bonds $\theta$ is 175$^{\circ}$ of 1$T''$ different from the 180$^{\circ}$ of 1T phase and the six Mo-S bond lengths of 1$T''$ don't have the same value. These small distortions can break the double degeneration of e$_{\rm g}$ orbitals (d$_{x^2-y^2}$, d$_{z^2}$) and produce the small bandgap [Fig.~\ref{fig:projectedband}(c)].
For the case of heavy chalcogens, X atom tinily moves backward the Mo atom, which strengthens the coupling between the X-p and Mo$_1$-d$_{z^2}$ orbitals and weakens coupling between the S-p and Mo$_1$-d$_{x^2-y^2}$ orbitals. These modulations of couplings lead to the higher bonding state of d$_{x^2-y^2}$ (CBM), the lower bonding state of d$_{z^2}$ (VBM) and the rise of $\Gamma_c$. Hence, the bigger bandgap exists in the cases of heavier chalcogens [Tab.~\ref{tab:bandgap}]. However, the space between W atomic layer and X atomic layer is smaller than that in MoX$_2$, so the effect of W atom contrary to that of heavy chalcogens and make WX$_2$ have very small bandgap even be metal.

In order to ensure the stability of 1$T''$ phase, we also calculate the phonon spectra. As shown in the bottom half of Fig.~\ref{fig:bandphonon}, only WSe$_2$ has the large imaginary frequency and other systems all have dynamics stabilities. Because the $\Gamma$ point has symmetry of C$_{3v}$ (3m) point group in 1$T''$ phase, 33 optical phonon modes can be decomposed by three irreducible representations: A$_1$ (8 modes), A$_2$ (3 modes) and E (11 double degenerate modes). With the increase of atomic mass, the highest phonon frequency obviously decrease, such as 448.6 cm$^{-1}$ of MoS$_2$ and 235.8 cm$^{-1}$ of WTe$_2$. And the greater proportion of chalcogens also give rise to the more obvious changes of phonon frequency with the different chalcogens. In addition, the small mass ratio of M and X atoms can close the frequency gap between acoustic phonons and optic phonons, as shown in Fig.~\ref{fig:bandphonon}.

\begin{figure}[ht!]
\centerline{\includegraphics[width=0.4\textwidth]{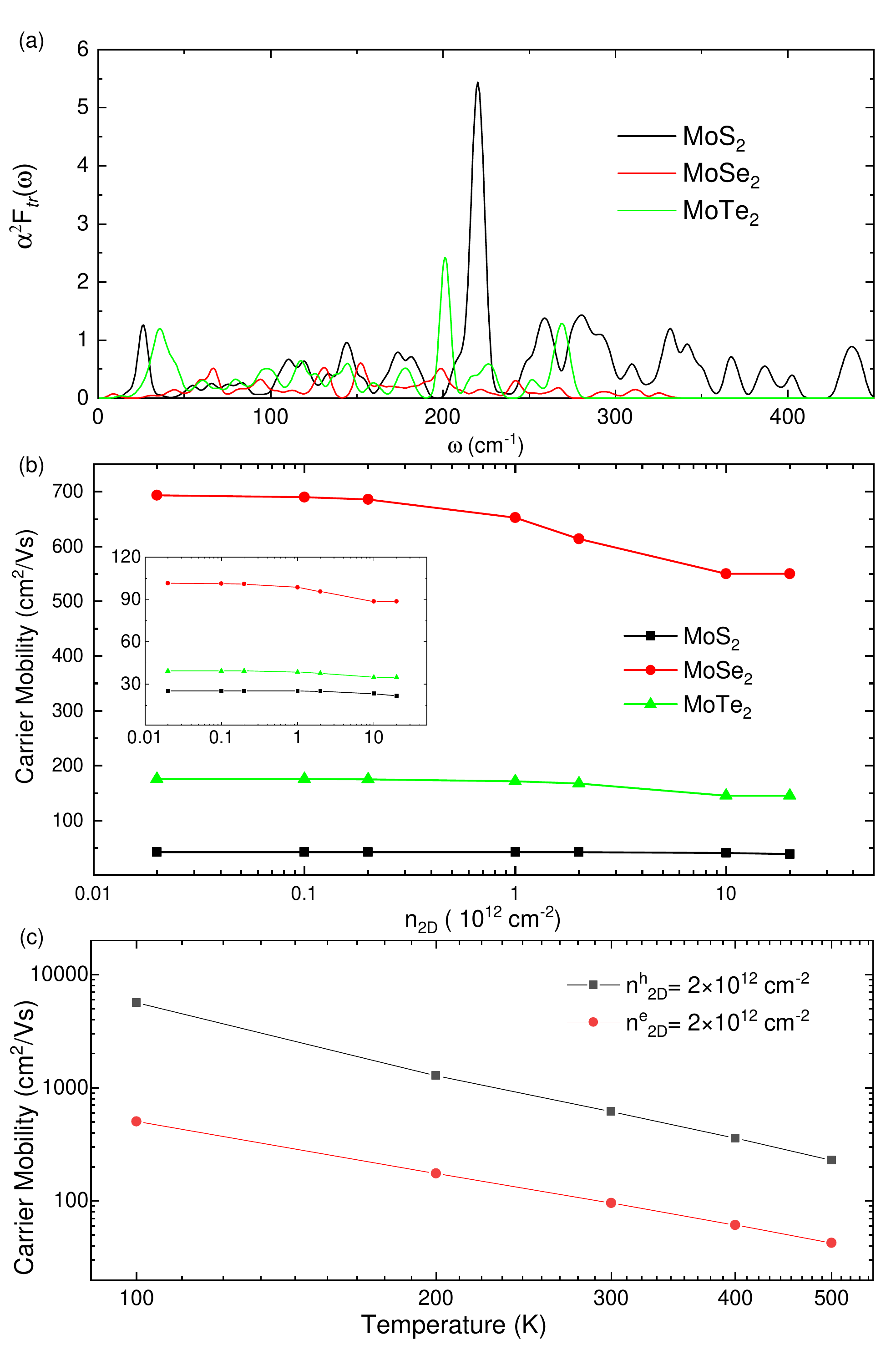}}
\caption{(a) Transport spectral function $\alpha_{tr}F(\omega)$ of MoS$_2$, MoSe$_2$ and MoTe$_2$ with hole carrier concentration n$^{\rm h}_{\rm 2D}$= 2$\times$10$^{12}$ cm$^{-2}$. (b) The hole carrier mobilities of MoS$_2$, MoSe$_2$ and MoTe$_2$ vary with the carrier concentration at room temperature. Inset: The electron carrier mobilities of MoS$_2$, MoSe$_2$ and MoTe$_2$ vary with the carrier concentration at room temperature. (c) The hole and electron carrier mobilities of MoSe$_2$ vary with the temperature when the carrier concentration n$_{\rm 2D}$= 2$\times$10$^{12}$ cm$^{-2}$.
\label{fig:mobility}}
\end{figure}

\begin{figure}[ht!]
\centerline{\includegraphics[width=0.4\textwidth]{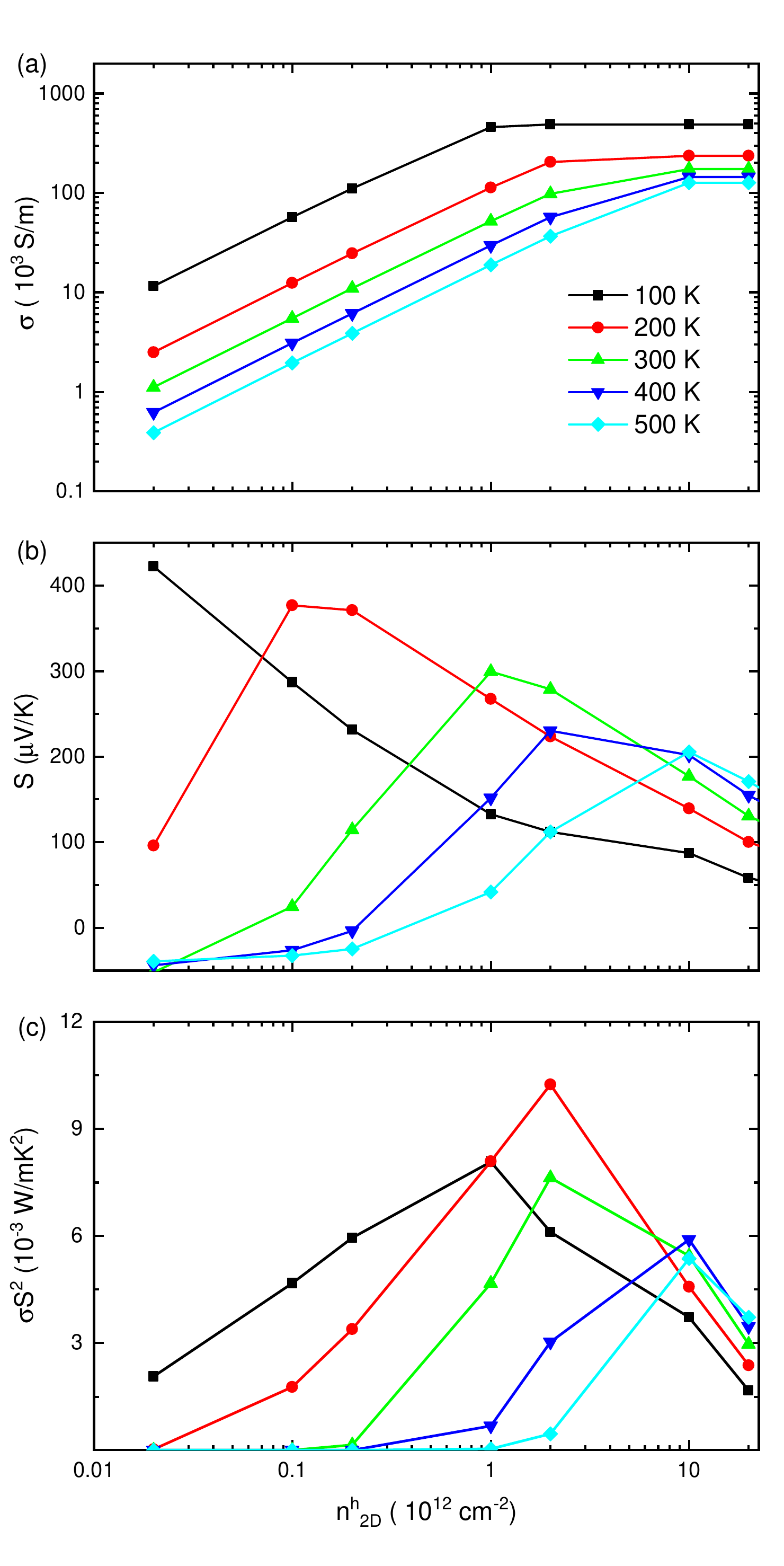}}
\caption{(a) Conductivity $\sigma$, (b) Seebeck coefficient S and (c) power factor $\sigma$S$^2$ vary with the hole carrier concentration at different temperatures. $\sigma$ is proportional to n$_{\rm 2D}$ and inversely proportional to T. The peak value of S inversely proportional to T and shift to high n$_{\rm 2D}$ with the increase of T.
\label{fig:powerfactor}}
\end{figure}

Basing on the stable semiconductor with suitable bandgap of MX$_2$ in $T''$ phase, we investigate the carrier doping and temperature dependences of mobility of MoX$_2$ (X=S, Se, Te) with the consideration of electron-phonon scattering. Firstly, we estimate the carrier effective mass of hole (m$^*_h$) and electron (m$^*_e$) on the basis of band structures and find that m$^*_h$ is lighter than m$^*_e$ of MoX$_2$ [Tab.~\ref{tab:bandgap}] and increase with atomic number of chalcogens.
by contrast, WTe$_2$ has a heavier m$^*_h$ than m$^*_e$, because VBM locate at $\Gamma$ point, differ from the K point for MoX$_2$.
Hence the next study keystone is hole-carrier transport properties and the doping range is set as 0.02 $\sim$ 20$\times$10$^{12}$ cm$^{-2}$. To facilitate analysis of relative contribution of phonons with different frequencies to electron-phonon scattering, we calculate the transport spectral function $\alpha^2_{tr}$F($\omega$)~\cite{Allen1978,Xu2013}, obtained by the phonon self-energy with doping in semiconductor. Figure~\ref{fig:mobility}(a) plots the $\alpha^2_{tr}$F($\omega$) of MoX$_2$ with n$^h_{\rm 2D}$=2$\times$10$^{12}$ cm$^{-2}$. It can be seen that the peak intensities of $\alpha^2_{tr}$F($\omega$) in MoS$_2$ are higher than those in other two cases and MoSe$_2$ has the lowest value in the whole spectrum space. Of particular note is the low frequency region around 40 cm$^{-1}$ and intermediate frequency region around 200 cm$^{-1}$. In the former, MoS$_2$ and MoTe$_2$ have strong electron-phonon coupling, which is almost absence from MoSe$_2$. And in the latter, the peak value of MoSe$_2$ is much smaller than that in MoS$_2$ or MoTe$_2$. From the above, MoSe$_2$ has the weakest electron-phonon coupling, to the benefit of high-performance carrier transport. As show in Fig.~\ref{fig:mobility}(b), the room-temperature hole carrier mobilities of MoS$_2$, MoSe$_2$ and MoTe$_2$ are 42, 690, and 176 cm$^2$/Vs at the low carrier concentration, respectively. It is noteworthy that the mobility of 1$T''$-MoSe$_2$ is much higher than that of 1H-phase TMDCs in experiments and predicted calculations~\cite{Radisavljevic2011,Levi2013,Kaasbjerg2012,Li2013,Li2015,Ge2014,Ong2013,Wang2016,Sohier2018}.
Here are two important factors need to be considered. One is the hole carrier effective mass, proportional to the atomic mass of chalcogens.
Other one is the electron-phonon coupling cause the scattering, whose the order of from weakest to strongest intensity is MoSe$_2<$ MoTe$_2<$ MoS$_2$. Thereby they result in the much higher hole carrier mobility of MoSe$_2$ than other two cases.
The down trend of the mobility on the carrier concentration also derive from the strong electron-phonon coupling of high density of electronic states at high concentration.
As a contrast, the electron carrier mobilities of MoX$_2$, plotted in the inset of Fig.~\ref{fig:mobility}(b), are lower than hole carrier by reason of heavy carrier effective mass.
Furthermore, phonon concentration has positive correlation relationship with temperature, thus the high temperature causes increased electron-phonon scattering, as the declining mobility of MoSe$_2$ with the increase of temperature [Fig.~\ref{fig:mobility}(c)].
And the temperature dependent hole carrier mobilities of MoX$_2$ (X=S, Se, Te) are fitted to be proportional to T$^{-2.0}$, T$^{-1.9}$ and T$^{-1.9}$, respectively.

Based the relaxation time of electron-phonon scattering, we calculate the electrical conductivity $\sigma$ and Seebeck coefficient S of MoSe$_2$ according to Eqs.(\ref{eq:sigma}) and (\ref{eq:seeb}). The proportionality between $\sigma$ and n$_{\rm 2D}$*$\mu$ can derive the ascending curve of $\sigma$ with carrier concentration and low $\sigma$ at high temperature, as shown in Fig.~\ref{fig:powerfactor}(a).
The hole-doping Seebeck coefficients as functions of carrier concentration at different temperatures are also plotted in Fig.~\ref{fig:powerfactor}(b). 1$T''$ MoSe$_2$ has a large Seebeck coefficient, and the maximum value of S, 422 $\mu$V/K of 100 K to 205 $\mu$V/K of 500 K,  shifts to high doping concentration and decrease as temperature increases, similar to the previous results of H-phase TMDCs~\cite{Kumar2015,Gandi2014}.
At room temperature, S can reach up to 300 $\mu$V/K when n$^h_{\rm 2D}$=1$\times10^{12}$ cm$^{-2}$, catching up to and even surpassing the experimental values of many two materials~\cite{GShi2015,Oh2016,Ng2019,Guo2018,Zeng2018,Hippalgaonkar2017,Yoshida2016,Pu2016,Saito2016}.
In Mott formula~\cite{Cutler1969}, hole-doping S of semiconductor is inversely proportional to doping concentration ( in direct proportion to the chemical potential).
However, the small bandgap easily causes the bipolar effect at low doping concentration~\cite{GShi2015}, which make S has proportional with doping concentration and the sign reversal of S with the increasing negative contribution of thermally excited electrons.
As shown in Fig.~\ref{fig:powerfactor}(c), the power factor ($\sigma$S$^2$) has a large value in the middle and low temperature zone (100 K$\sim$500 K). The highest value of 10.2$\times$10$^{-3}$ W/mK$^2$ with n$^h_{\rm 2D}$=2$\times10^{12}$ cm$^{-2}$@200 K. And it is more important that over a large temperature range, the maximal power factor of different temperatures can stay around $\sim$6.0 10$^{-3}$ W/mK$^2$, well above the present the experimental measurements of intrinsic power factor in the vast majority of TMDCs and some classic TE materials, such as SnSe, Bi$_2$Te$_3$ and PbTe~\cite{Ng2019,Heremans2008,Oh2016,Zhao2016}. It indicates the 1$T''$-phase MoSe$_2$ as a high-performance candidate TE materials in the low to intermediate temperature range.

\begin{figure}[ht!]
\centerline{\includegraphics[width=0.5\textwidth]{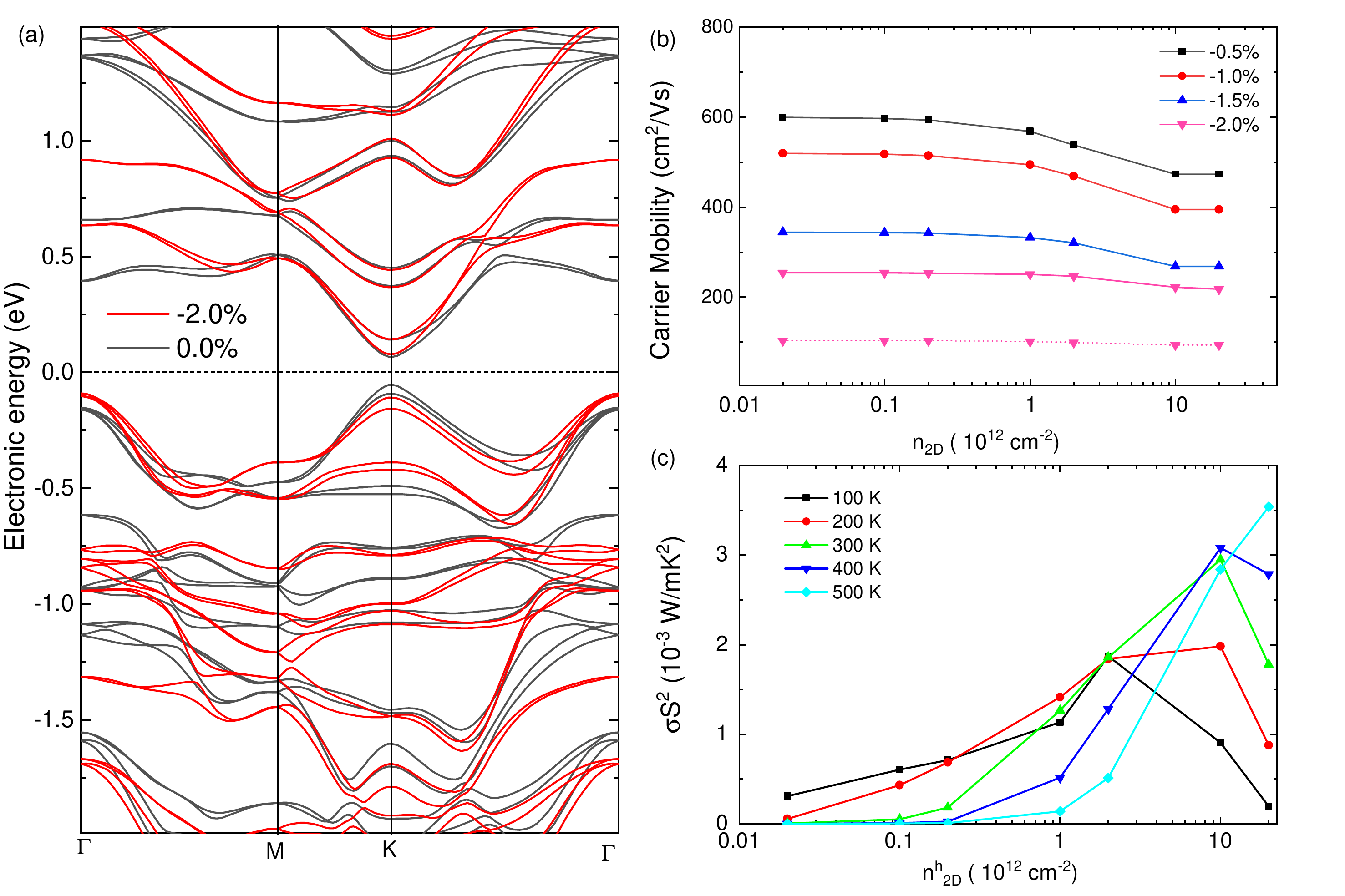}}
\caption{(a) Band structure of MoSe$_2$ under the compressive strain of $\epsilon$=-2.0\%. (b) The hole carrier mobilities of MoSe$_2$ vary with the carrier concentration under four compressive strains (-0.5\%, -1.0\%, -1.5\% and -2\%). (c) power factor $\sigma$S$^2$ vary with the hole carrier concentration under compressive strain of $\epsilon$=-2.0\%.
\label{fig:strain}}
\end{figure}

Inspired by the similar effect of heavy chalcogens and compressive strain, which both lead to the increase of monolayer thickness and impact the interaction between Mo and Se atoms~\cite{Chang2013}, we also expect and study the effect of small compressive strain ( $\epsilon$=($a-a_0$)/$a_0\times$100\%$\leq$-2.0\%) on the transport property of MoSe$_2$, in order to enhance the bandgap as well as the temperature range of thermoelectric application. In the electronic band structure with $\epsilon$=-2.0\%, the bandgap increases to 0.19 eV and the hole carrier effective mass m$^*_h$ is light to 0.488 m$_0$ [Fig.~\ref{fig:strain}]. But the energy difference between $\Gamma_v$ and VBM almost disappear, which can enhance the intervalley scattering of $\Gamma$ and K, assisted by the K-vector phonons, similar to the intervalley scattering in 1H MoS$_2$~\cite{Li2013,Ge2014}. Consequently, after the introduction of small compressive strain, the hole carrier mobility drops as well as the the decrease of power factor (1$\sim$3 10$^{-3}$ W/mK$^2$) [Fig.~\ref{fig:strain}]. And the peak values of power factor all locate at the range of high concentration ($\geq 2 $ 10$^{12}$ cm$^{-2}$ ).

\section{Conclusion}

In summary, by using the first-principles calculations with Boltzmann transport theory, we studies systematically the metastable monolayer 1$T''$ phase MX$_2$, including electronic structure, electron-phonon coupling, carrier mobility and TE power factor. The small direct bandgap at K point of three molybdenum compounds is attributed to the distorted octahedral coordination of [MoX$_6$]. And the extremely weak electron-phonon coupling of MoSe$_2$ gives rise to its hole carrier mobility as high as 690 cm$^2$/Vs at 300K. Moreover, combining the Seebeck coefficient around 300 $\mu$V/K, it is obtained that the TE power factor of MoSe$_2$ keeps above 6 10$^{-3}$W/mK$^2$ in the large range of temperature (100K$\sim$500K). Our results illustrate the outstanding potential 1$T''$ MoSe$_2$ on TE materials.

\begin{acknowledgments}
This work was supported by the NSFC (Grants No.11747054), the Specialized Research Fund for the Doctoral Program of Higher Education of China (Grant No.2018M631760), the Project of Heibei Educational Department, China (No. ZD2018015 and QN2018012), and the Advanced Postdoctoral Programs of Hebei Province (No.B2017003004).
\end{acknowledgments}


\begin{thebibliography}{100}
\expandafter\ifx\csname url\endcsname\relax
  \def\url#1{\texttt{#1}}\fi
\expandafter\ifx\csname urlprefix\endcsname\relax\def\urlprefix{URL }\fi
\providecommand{\bibinfo}[2]{#2}





\bibitem{Feng2012}J. Feng, X. Qian, C.-W. Huang, and J. Li, Strain-engineered artificial atom as a broad-spectrum solar energy funnel, Nat. Photon. 6, 866 (2012).
\bibitem{Conley2013}H. J. Conley, B. Wang, J. I. Ziegler, R. F. Haglund, S. T. Pantelides, and K. I. Bolotin, Bandgap Engineering of Strained Monolayer and Bilayer MoS2, Nano Lett. 13, 3626 (2013).
\bibitem{Castellanos-Gomez2013}A. Castellanos-Gomez, R. Roldan, E. Cappelluti, M. Buscema, F. Guinea, H. S. J. van der Zant, and G. A. Steele, Local Strain Engineering in Atomically Thin MoS2, Nano Lett. 13, 5361 (2013).
\bibitem{Chhowalla2013}M. Chhowalla, H. S. Shin, G. Eda, L. J. Li, K. P. Loh, and H. Zhang, The chemistry of two-dimensional layered transition metal dichalcogenide nanosheets, Nat. Chem. 5, 263 (2013).
\bibitem{Wang2012}Q. H. Wang, K. Kalantar-Zadeh, A. Kis, J. N. Coleman, and M. S. Strano, Electronics and optoelectronics of two-dimensional transition metal dichalcogenides, Nat. Nanotechnol. 7, 699 (2012).
\bibitem{Fiori2014}G. Fiori, F. Bonaccorso, G. Iannccone, T. Palacios, D. Neumaier, A. Seabaugh, S. K. Banerjee, and L. Colombo, Electronics based on two-dimensional materials, Nat. Nanotechnol. 9, 768 (2014).
\bibitem{Bernardi2017}M. Bernardi, C. Ataca, M. Palummo, and J. C. Grossman, Optical and electronic properties of two-dimensional layered materials, Nanophotonics 6, 479 (2017).
\bibitem{Britnell2013}L. Britnell, K. S. Novoselov, R. M. Ribiro, A. Eckmann, R. Jalil, A. Mishchenko, Y. J. Kim, R. V. Gorbachev, T. Georgiou, S. V. Morozov, A. N. Grigorenko, A. K. Geim, C. Casiraghi, A. H. Castro Neto, and K. S. Novoselov, Strong light-matter interactions in heterostructures of atomically thin films, Science 340, 1311 (2013).
\bibitem{Fang2012}H. Fang, S. Chuang, T. C. Chang, K. Takei, T. Takahashi, and A. Javey, High-performance single layered WSe2 p-FETs with chemically doped contacts, Nano Lett. 12, 3788 (2012).
\bibitem{Liu2013}W. Liu, J. Kang, D. Sarkar, Y. Khatami, D. Jena, and K. Banerjee, Role of metal contacts in designing high-performance monolayer n-type WSe2 field effect transistors, Nano Lett. 13, 1983 (2013).
\bibitem{Yoon2011}Y. Yoon, K. Ganapathi, and S. Salahuddin, How Good Can Monolayer MoS2 Transistors Be, Nano Lett. 11, 3768 (2011).
\bibitem{Mak2013}K. F. Mak, K. He, C. Lee, G. H. Lee, J. Hone, T. F. Heinz, and J. Shan, Tightly bound trions in monolayer MoS2, Nat. Mater. 12, 207 (2013).
\bibitem{Qiu2015}D. Y. Qiu, T. Cao, and S. G. Louie, Nonanalyticity, valley quantum phases, and lightlike exciton dispersion in monolayer transition metal dichalcogenides: Theory and first-principles calculations, Phys. Rev. Lett. 115, 176801 (2015).
\bibitem{Huo2014}N. Huo, J. Kang, Z. Wei, S.-S. Li, J. Li, and S.-H. Wei, Novel and Enhanced Optoelectronic Performances of Multilayer MoS2-WS2 Heterostructure Transistors, Adv. Funct. Mater. 24, 7025 (2014).
\bibitem{Radisavljevic2011}B. Radisavljevic, A. Radenovic, J. Brivio, V. Giacometti, and A. Kis, Single-layer MoS2 transistors, Nat. Nanotechnol. 6, 147 (2011).
\bibitem{Levi2013}R. Levi, O. Bitton, G. Leitus, R. Tenne, and E. Joselevich, Field-Effect Transistors Based on WS2 Nanotubes with High Current-Carrying Capacity, Nano Lett. 13, 3736 (2013).
\bibitem{Xu2014}X. D. Xu, W. Yao, D. Xiao, and T. F. Heinz, Spin and pseudospins in layered transition metal dichalcogenides, Nat. Phys. 10, 343 (2014).
\bibitem{Mak2014}K. F. Mak, K. L. McGill, J. Park, and P. L. McEuen, The valley Hall effect in MoS2 transistors, Science 344, 1489 (2014).
\bibitem{Cui2015}X. Cui, G. H. Lee, Y. D. Kim, G. Arefe, P. Y. Huang, C. H. Lee, D. A. Chenet, X. Zhang, L. Wang, F. Ye, F. Pizzocchero, B. S. Jessen, K. Watanabe, T. Taniguchi, D. A. Muller, T. Low, P. Kim, and J. Hone, Multi-terminal transport measurements of MoS2 using a van der Waals heterostructure device platform, Nat. Nanotechnol. 10, 534 (2015).
\bibitem{Roy2013}K. Roy, M. Padmanabhan, S. Goswami, T. P. Sai, G. Ramalingam, S. Raghavan, and A. Ghosh, Graphene-MoS2 hybrid structures for multifunctional photoresponsive memory devices, Nat. Nanotechnol. 8, 826 (2013).
\bibitem{Qi2015}J. J. Qi, Y.W. Lan, A. Z. Stieg, J. H. Chen, Y. L. Zhong, L. J. Li, C. D. Chen, Y. Zhang, and K. L. Wang, Piezoelectric effect in chemical vapour deposition-grown atomic-monolayer triangular molybdenum disulfide piezotronics, Nat. Commun. 6, 7430 (2015).
\bibitem{Song2015}I. Song, C. Park, and H. C. Choi, Synthesis and properties of molybdenum disulphide: from bulk to atomic layers, RSC Adv. 5, 7495 (2015).
\bibitem{Bruyer2016}E. Bruyer, D. Di Sante, P. Barone, A. Stroppa, M. H. Whangbo, and S. Picozzi, Possibility of combining ferroelectricity and Rashba-like spin splitting in monolayers of the 1T-type transition-metal dichalcogenides MX2(M=Mo,W;X=S,Se,Te), Phys. Rev. B 94, 195402 (2016).
\bibitem{Zhang2018}X. Zhang, Z. Lai, Q. Ma, and H. Zhang, Novel structured transition metal dichalcogenide nanosheets, Chem. Soc. Rev. 47, 3301 (2018).
\bibitem{Linghu2019}Y. Linghu, N. Li, Y. Du, and C. Wu, Ligand induced structure and property changes of 1T-MoS2, Phys. Chem. Chem. Phys. 21, 9391 (2019).
\bibitem{Zhao2018} W. Zhao, J. Pan, Y. Fang, X. Che, D. Wang, K. Bu, and F. Huang, Metastable MoS2 Crystal Structure, Electronic Band Structure, Synthetic Approach and Intriguing Physical Properties, Chem. Eur. J. 24, 15942 (2018).
\bibitem{Kan2014}M. Kan, J. Y. Wang, X. W. Li, S. H. Zhang, Y. W. Li, Y. Kawazoe, Q. Sun, and P. Jena, Structures and Phase Transition of a MoS2 Monolayer, J. Phys. Chem. C 118, 1515 (2014).
\bibitem{Calandra2013} M. Calandra, Chemically exfoliated single-layer MoS2: Stability, lattice dynamics, and catalytic adsorption from first principles, Phys. Rev. B 88, 245428 (2013).
\bibitem{Zhuang2017}H. L. Zhuang, M. D. Johannes, A. K. Singh, and R. G. Hennig, Doping-controlled phase transitions in single-layer MoS2, Phys. Rev. B 96, 165305 (2017).
\bibitem{Pal2017}B. Pal, A. Singh, G. Sharada, P. Mahale, A. Kumar, S. Thirupathaiah, H. Sezen, M. Amati, L. Gregoratti, U. V. Waghmare, and D. D. Sarma, Chemically exfoliated MoS2 layers: Spectroscopic evidence for the semiconducting nature of the dominant trigonal metastable phase, Phys. Rev. B 96, 195426 (2017).
\bibitem{Zhou2018}J. Zhou, J. Lin, X. Huang, Y. Zhou, Y. Chen, J. Xia, H. Wang, Y. Xie, H. Yu, J. Lei, D. Wu, F. Liu, Q. Fu, Q. Zeng, C.-H. Hsu, C. Yang, L. Lu, T. Yu, Z. Shen, H. Lin, B. I. Yakobson, Q. Liu, K. Suenaga, G. Liu, and Z. Liu, A library of atomically thin metal chalcogenides, Nature 556, 355 (2018).
\bibitem{Singh2015}A. Singh, S. N. Shirodkar, and U. V. Waghmare, 1H and 1T polymorphs, structural transitions and anomalous properties of (Mo, W)(S, Se)2 monolayers: first-principles analysis, 2D Materials 2, 035013 (2015).
\bibitem{Chou2015}S. S. Chou, N. Sai, P. Lu, E. N. Coker, S. Liu, K. Artyushkova, T. S. Luk, B. Kaehr, and C. J. Brinker, Understanding catalysis in a multiphasic two-dimensional transition metal dichalcogenide, Nat. Commun. 6, 8311 (2015).
\bibitem{Eda2012}G. Eda, T. Fujita, H. Yamaguchi, D. Voiry, M. Chen, and M. Chhowalla, Coherent Atomic and Electronic Heterostructures of Single-Layer MoS2, ACS Nano 6, 7311 (2012).
\bibitem{Yu2018}Y. Yu, G. H. Nam, Q. He, X. J. Wu, K. Zhang, Z. Yang, J. Chen, Q. Ma, M. Zhao, Z. Liu, F.-R. Ran, X. Wang, H. Li, X. Huang, B. Li, Q. Xiong, Q. Zhang, Z. Liu, L. Gu, Y. Du, W. Huang, and H. Zhang, High phase-purity 1T¡ä-MoS 2-and 1T¡ä-MoSe 2-layered crystals, Nat. Chem. 10, 638 (2018).
\bibitem{Fang2018}Y. Fang, X. Hu, W. Zhao, J. Pan, D. Wang, K. Bu, Y. Mao, S. Chu, P. Liu, T. Zhai, and F. Huang, Structural Determination and Nonlinear Optical Properties of New 1T'-Type MoS2 Compound, J. Am. Chem. Soc. 141, 790 (2019).
\bibitem{Shirodkar2014}S. N. Shirodkar, and U. V. Waghmare, Emergence of Ferroelectricity at a Metal-Semiconductor Transition in a
1T Monolayer of MoS2, Phys. Rev. Lett. 112, 157601 (2014).
\bibitem{Shang2018}C. Shang, Y. Q. Fang, Q. Zhang, N. Z. Wang, Y. F. Wang, Z. Liu, B. Lei, F. B. Meng, L. K. Ma, T. Wu, Z. F. Wang, C. G. Zeng, F. Q. Huang, Z. Sun, and X. H. Chen, Superconductivity in the metastable 1T' and 1T''' phases of MoS2 crystals, Phys. Rev. B 98, 184513 (2018).
\bibitem{Acerce2015}M. Acerce, D. Voiry, and M. Chhowalla, Metallic 1T phase MoS2 nanosheets as supercapacitor electrode materials, Nat. Nanotechnol. 10, 313 (2015).
\bibitem{Voiry2013}D. Voiry, M. Salehi, R. Silva, T. Fujita,M. Chen, T. Asefa, V. B. Shenoy, G. Eda, and M. Chhowalla, Conducting MoS2 nanosheets as catalysts for hydrogen evolution reaction, Nano Lett. 13, 6222 (2013).
\bibitem{Lukowski2013}M. A. Lukowski, A. S. Daniel, F. Meng, A. Forticaux, L. Li, and S. Jin, Enhanced Hydrogen Evolution Catalysis from Chemically Exfoliated Metallic MoS2 Nanosheets, J. Am. Chem. Soc. 135, 10274 (2013).
\bibitem{Qian2014}X. F. Qian, J. W. Liu, L. Fu, and J. Li, Quantum spin Hall effect in two-dimensional transition metal, Science 346, 1344 (2014).
\bibitem{Choe2016}D.-H. Choe, H.-J. Sung, and K. J. Chang, Understanding topological phase transition in monolayer transition metal dichalcogenides, Phys. Rev. B 93, 125109 (2016).
\bibitem{Tang2017}S. Tang, C. Zhang, D. Wong, Z. Pedramrazi, H.-Z. Tsai, C. Jia, B. Moritz, M. Claassen, H. Ryu, S. Kahn, J. Jiang, H. Yan, M. Hashimoto, D. Lu, R. G. Moore, C.-C. Hwang, C. Hwang, Z. Hussain, Y. Chen, M. M. Ugeda, Z. Liu, X. Xie, T. P. Devereaux, M. F. Crommie, S.-K. Mo, and Z.-X. Shen, Quantum spin Hall state in monolayer 1T'-WTe2, Nat. Phys. 13, 683 (2017).
\bibitem{Cutler1969}M. Cutler, and N. Mott, Observation of Anderson Localization in an Electron Gas, Phys. Rev. 181, 1336 (1969).
\bibitem{Bell2008}1. L. E. Bell, Cooling, Heating, Generating Power, and Recovering Waste Heat with Thermoelectric Systems, Science 321, 1457 (2008).
\bibitem{Heremans2013}J. P. Heremans, M. S. Dresselhaus, L. E. Bell, and D. T. Morelli, When thermoelectrics reached the nanoscale, Nat. Nanotechnol. 8, 471 (2013).
\bibitem{Dresselhaus2007}M. S. Dresselhaus, G. Chen, M. Y. Tang, R. G. Yang, H. Lee, D. Z. Wang, Z. F. Ren, J.-P. Fleurial, and P. Gogna, New directions for low-dimensional thermoelectric materials, Adv. Mater. 19, 1043 (2007).
\bibitem{Zhao2014}L. D. Zhao, V. P. Dravid, and M. G. Kanatzidis, The panoscopic approach to high performance thermoelectrics, Energy Environ. Sci. 7, 251 (2014).
\bibitem{Huang2013}W. Huang, H. Da, and G. Liang, Thermoelectric performance of MX2 (M=Mo,W; X=S,Se) monolayers, J. Appl. Phys. 113, 104304 (2013).
\bibitem{Huang2014}W. Huang, X. Luo, C. K. Gan, S. K. Quek, and G. Liang, Theoretical study of thermoelectric properties of few-layer MoS2 and WSe2, Phys. Chem. Chem. Phys. 16, 10866 (2014).
\bibitem{Babaei2014}H. Babaei, J. M. Khodadadi, and S. Sinha, Large theoretical thermoelectric power factor of suspended single-layer MoS2, Appl. Phys. Lett. 105, 193901 (2014).
\bibitem{Wang2017}R. N. Wang, G. Y. Dong, S. F. Wang, G. S. Fu, and J. L. Wang, Variations of thermoelectric performance by electric fields in bilayer MX2 (M=W, Mo; X=S, Se), Phys. Chem. Chem. Phys. 19, 5797 (2017).
\bibitem{Wu2014}J. Wu, H. Schmidt, K. K. Amara, X. Xu, G. Eda, and B. Ozyilmaz, Large Thermoelectricity via Variable Range Hopping in Chemical Vapor Deposition Grown Single-Layer MoS2, Nano Lett. 14, 2730 (2014).
\bibitem{Yoshida2016}M. Yoshida, T. Iizuka, Y. Saito, M. Onga, R. Suzuki, Y. Zhang, Y. Iwasa, and S. Shimizu, Gate-Optimized Thermoelectric Power Factor in Ultrathin WSe2 Single Crystals, Nano Lett. 16, 2061 (2016).
\bibitem{TWang2016}T. Wang, C. Liu, J. Xu, Z. Zhu, E. Liu, Y. Hu, C. Li, and F. Jiang, Thermoelectric performance of restacked MoS2 nanosheets thin-film, Nanotechnology 27, 285703 (2016).
\bibitem{Fan2014}D. D. Fan, H. J. Liu, L. Cheng, P. H. Jiang, J. Shi, and X. F. Tang, MoS2 nanoribbons as promising thermoelectric materials, Appl. Phys. Lett., 105, 133113 (2014).
\bibitem{Hippalgaonkar2017}K. Hippalgaonkar, Y. Wang, Y. Ye, D. Y. Qiu, H. Zhu, Y. Wang, J. Moore, S. G. Louie, and X. Zhang,
High thermoelectric power factor in two-dimensional crystals of MoS2, Phys. Rev. B 95, 115407 (2017).
\bibitem{Wu2019}J. Wu, Y. Liu, Y. Liu, Y. Cai, Y. Zhao, H. K. Ng, K. Watanabe, T. Taniguchi, G. Zhang, C. Qiu, D. Chi, A. C. Neto, J. T. L. Thong, K. P. Loh, and K. Hippalgaonkar, Kondo Impurities in Two Dimensional MoS2 for Achieving Ultrahigh Thermoelectric Powerfactor, arXiv:1901.04661 (2019).
\bibitem{Gascoin2005}F. Gascoin, S. Ottensmann, D. Stark, S. Haile, and G. Snyder, Zintl Phases as Thermoelectric Materials Tuned Transport Properties of the Compounds CaxYb1¨CxZn2Sb2, Adv. Funct. Mater. 15, 1860 (2005).
\bibitem{Lee2012}M.-S. Lee, and S. D. Mahanti, Validity of the rigid band approximation in the study of the thermopower of narrow band gap semiconductors, Phys. Rev. B 85, 165149 (2012).
\bibitem{Goldsmid2014}H. J. Goldsmid, Bismuth Telluride and Its Alloys as Materials for Thermoelectric Generation, Materials 7, 2577 (2014).
\bibitem{Giustino2017}F. Giustino, Electron-phonon interactions from first principles, Rev. Mod. Phys. 89, 015003 (2017).
\bibitem{Ponce2018}S. Ponce, E. R. Margine, and F. Giustino, Towards predictive many-body calculations of phonon-limited carrier mobilities in semiconductors, Phys. Rev. B 97, 121201(R) (2018).
\bibitem{Baroni2001}S. Baroni, S. de Gironcoli, A. Dal Corso, and P. Giannozzi, Phonons and related crystal properties from density-functional perturbation theory, Rev. Mod. Phys. 73, 515 (2001).
\bibitem{Huang2018}S. Huang, H. J. Liu, D. D. Fan, P. H. Jiang, J. H. Liang, G. H. Cao, R. Z. Liang, and J. Shi, First-Principles Study of the Thermoelectric Properties of the Zintl Compound KSnSb, J. Phys. Chem. C 122, 4217 (2018).
\bibitem{Pizzi2014}G. Pizzi, D. Volja, B. Kozinsky, M. Fornari, and N. Marzari, BoltzWann: A code for the evaluation of thermoelectric and electronic transport properties with a maximally-localized Wannier functions basis, Comput. Phys. Commun. 185, 422 (2014).
\bibitem{Giannozzi2017}P. Giannozzi, O. Andreussi, T. Brumme, O. Bunau, M. B. Nardelli, M. Calandra, R. Car, C. Cavazzoni, D. Ceresoli, M.
Cococcioni, N. Colonna, I. Carnimeo, A. D. Corso, S. de Gironcoli, P. Delugas, R. DiStasio, A. Ferretti, A. Floris, G. Fratesi, G. Fugallo, R. Gebauer, U. Gerstmann., F. Giustino, T. Gorni, J. Jia, M. Kawamura, H.-Y. Ko, A. Kokalj, E. Ku?ukbenli, M. Lazzeri, M. Marsili, N. Marzari, F. Mauri, N. L. Nguyen, H.-V. Nguyen, A. Otero-de-la-Roza, L. Paulatto, S. Ponce, D. Rocca, R. Sabatini, B. Santra, M. Schlipf, A. P. Seitsonen, A. Smogunov, I. Timrov, T. Thonhauser, P. Umari, N. Vast, X. Wu, and S. Baroni, Advanced capabilities for materials modelling with Quantum ESPRESSO, J. Phys.: Condens. Matter 29, 465901 (2017).
\bibitem{Perdew1996}J. P. Perdew, K. Burke, and M. Ernzerhof, Generalized Gradient Approximation Made Simple, Phys. Rev. Lett. 77, 3865 (1996).
\bibitem{Ponce2016}S. Ponce, E. R. Margine, C. Verdi, and F. Giustino, EPW: Electron-phonon coupling, transport and superconducting properties using maximally localized Wannier functions, Comput. Phys. Commun. 209, 116 (2016).
\bibitem{Mostofi2014}A. A. Mostofi, J. R. Yates, G. Pizzi, Y.-S. Lee, I. Souza, D. Vanderbilt, and N. Marzari, An updated version of wannier90: A tool for obtaining maximally-localised Wannier functions, Comput. Phys. Commun. 185, 2309 (2014).
\bibitem{Mostofi2008}A. A. Mostofi, J. R. Yates, Y.-S. Lee, I. Souza, D. Vanderbilt, and N. Marzari, wannier90: A tool for obtaining maximally-localised Wannier functions, Comput. Phys. Commun. 178, 685 (2008).
\bibitem{Giustino2007}F. Giustino, M. L. Cohen, and S. G. Louie, Electron-phonon interaction using Wannier functions, Phys. Rev. B 76, 165108 (2007).
\bibitem{Marzari2012}N. Marzari, A. A. Mostofi, J. R. Yates, I. Souza, and D. Vanderbilt, Maximally localized Wannier functions: Theory and applications, Rev. Mod. Phys. 84, 1419 (2012).
\bibitem{Allen1978}P. B. Allen, New method for solving Boltzmann's equation for electrons in metals, Phys. Rev. B 17, 3725 (1978).
\bibitem{Xu2013}B. Xu, and M. J. Verstraete, First-principles study of transport properties in Os and OsSi, Phys. Rev. B 87, 134302 2013).
\bibitem{Kaasbjerg2012}K. Kaasbjerg, K. S. Thygesen, and K. W. Jacobsen, Phonon-limited mobility in n-type single-layer MoS2 from first principles, Phys. Rev. B 85, 115317 (2012).
\bibitem{Li2013}X. Li, J. T. Mullen, Z. Jin, K. M. Borysenko, M. Buongiorno Nardelli, and K. W. Kim, Intrinsic electrical transport properties of monolayer silicene and MoS2 from first principles, Phys. Rev. B 87, 115418 (2013).
\bibitem{Li2015}W. Li, Electrical transport limited by electron-phonon coupling from Boltzmann transport equation: An ab initio study of Si, Al, and MoS2, Phys. Rev. B 92, 075405 (2015).
\bibitem{Ge2014}Y. Ge, W. Wan, W. Feng, D. Xiao, and Y. Yao, Effect of doping and strain modulations on electron transport in monolayer MoS2, Phys. Rev. B 90, 035414 (2014).
\bibitem{Ong2013}Z.-Y. Ong, and M. V. Fischetti, Mobility enhancement and temperature dependence in top-gated single-layer MoS2, Phys. Rev. B 88, 165316 (2013).
\bibitem{Wang2016} J. Wang, Q. Yao, C. Huang, X. Zou, L. Liao, S. Chen, Z. Fan, K. Zhang, W. Wu, X. Xiao, C. Jiang, and W. Wu, High Mobility MoS2 Transistor with Low Schottky Barrier Contact by Using Atomic Thick h-BN as a Tunneling Layer, Adv. Mater. 28, 8302 (2016).
\bibitem{Sohier2018} T. Sohier, D. Campi, N. Marzari, and M. Gibertini, Mobility of two-dimensional materials from first principles in an accurate and automated framework, Phys. Rev. Materials 2, 114010 (2018).
\bibitem{Kumar2015}S. Kumar, and U. Schwingenschlogl, Thermoelectric Response of Bulk and Monolayer MoSe2 and WSe2, Chem. Mater. 27, 1278 (2015).
\bibitem{Gandi2014}A. N. Gandi, and U. Schwingenschlogl, WS2 As an Excellent High-Temperature Thermoelectric Material, Chem. Mater. 26, 6628 (2014).
\bibitem{GShi2015}G. Shi, and E. Kioupakis, Quasiparticle band structures and thermoelectric transport properties of p-type SnSe, J. Appl. Phys. 117, 065103 (2015).
\bibitem{Guo2018}Y. Guo, C. Dun, J. Xu, P. Li, W. Huang, J. Mu, C. Hou, C. A. Hewitt, Q. Zhang, Y. Li, D. L. Carroll, and H. Wang, Wearable Thermoelectric Devices Based on Au-Decorated Two-Dimensional MoS2, ACS Applied Materials \& Interfaces 10, 33316 (2018).
\bibitem{Zeng2018}J. Zeng, X. He, S.-J. Liang, E. Liu, Y. Sun, C. Pan, Y. Wang, T. Cao, X. Liu, C. Wang, L. Zhang, S. Yan, G. Su, Z. Wang, K. Watanabe, T. Taniguchi, D. J. Singh, L. Zhang, and F. Miao, Experimental Identification of Critical Condition for Drastically Enhancing Thermoelectric Power Factor of Two-Dimensional Layered Materials, Nano Lett. 18, 7538 (2018).
\bibitem{Pu2016}J. Pu, K. Kanahashi, N. T. Cuong, C.-H. Chen, L.-J. Li, S. Okada, H. Ohta, and T. Takenobu, Enhanced thermoelectric power in two-dimensional transition metal dichalcogenide monolayers, Phys. Rev. B 94, 014312 (2016).
\bibitem{Saito2016}Y. Saito, T. Iizuka, T. Koretsune, R. Arita, S. Shimizu, and Y. Iwasa, Gate-Tuned Thermoelectric Power in Black Phosphorus, Nano Lett. 16, 4819 (2016).
\bibitem{Oh2016}J. Young Oh, J. H. Lee, S. Woong Han, S. Chae, E. Bae, Y. Kang, W. J. Choi, S. Y. Cho, J.-O. Lee, B. H. Koo, and T. T. Lee, Chemically exfoliated transition metal dichalcogenide nanosheet-based wearable thermoelectric generators. Energy Environ. Sci. 9. 1696 (2016).
\bibitem{Ng2019}H. Kuan Ng, A. Abutaha, D. Voiry, I. Verzhbitskiy, Y. Cai, G. Zhang, Y. Liu, J. Wu, M. Chhowalla, G. Eda, and K. Hippalgaonkar, Effects Of Structural Phase Transition On Thermoelectric Performance in Lithium-Intercalated Molybdenum Disulfide (LixMoS2), ACS Applied Materials \& Interfaces 11, 12184 (2019).
\bibitem{Heremans2008}J. P. Heremans, V. Jovovic, E. S. Toberer, A. Saramat, K. Kurosaki, A. Charoenphakdee, S. Yamanaka, and G. J. Snyder, Enhancement of thermoelectric efficiency in PbTe by distortion of the electronic density of states, Science 321, 554 (2008).
\bibitem{Zhao2016}L.-D. Zhao, G. Tan, S. Hao, J. He, Y. Pei, H. Chi, H. Wang, S. Gong, H. Xu, V. P. Dravid, C. Uher, G. J. Snyder, C. Wolverton, and M. G. Kanatzidis, Ultrahigh power factor and thermoelectric performance in hole-doped single-crystal SnSe, Science 351, 141 (2016).
\bibitem{Chang2013}C. H. Chang, X. F. Fan, S. H. Lin and J. L. Kuo, Orbital Analysis of Electronic Structure and Phonon Dispersion in MoS2, MoSe2, WS2, and WSe2 Monolayers under Strain, Phys. Rev. B 88, 195420 (2013).












\end{thebibliography}
\end{document}